\documentclass[aps,prb, twocolumn, superscriptaddress, floatfix,10pt]{revtex4}
\usepackage{amssymb}
\usepackage{amsmath}
\usepackage{graphicx}
\usepackage{color}
\begin{document}


\title{Hyperpolarized Long-\emph T$\bf _1$ Silicon Nanoparticles for Magnetic Resonance Imaging}

\author{J.~W.~Aptekar}
\altaffiliation{These authors contributed equally to this work}
\affiliation{Department of Physics, Harvard University, Cambridge, Massachusetts 02138, USA}
\author{M.~C.~Cassidy}
\altaffiliation{These authors contributed equally to this work}
\affiliation{Department of Physics, Harvard University, Cambridge, Massachusetts 02138, USA}
\author{A.~C.~Johnson}
\affiliation{Department of Physics, Harvard University, Cambridge, Massachusetts 02138, USA}
\author{R.~A.~Barton}
\affiliation{Department of Physics, Harvard University, Cambridge, Massachusetts 02138, USA}
\author{M.~Y.~Lee}
\affiliation{Department of Physics, Harvard University, Cambridge, Massachusetts 02138, USA}
\author{A.~C.~Ogier}
\affiliation{Department of Physics, Harvard University, Cambridge, Massachusetts 02138, USA}
\author{C.~Vo}
\affiliation{Department of Physics, Harvard University, Cambridge, Massachusetts 02138, USA}
\author{M.~N.~Anahtar}
\affiliation{}
\author{Y.~Ren}
\affiliation{}
\author{S.~N.~Bhatia}
\affiliation{Harvard-MIT Division of Health Sciences and Technology, Massachusetts Institute of Technology E19-502D Cambridge, MA 02139, USA}
\author{C.~Ramanathan}
\affiliation{Department of Nuclear Science and Engineering, Massachusetts Institute of Technology, Cambridge, MA 02139, USA}
\author{D.~G.~Cory}
\affiliation{Department of Nuclear Science and Engineering, Massachusetts Institute of Technology, Cambridge, MA 02139, USA}
\author{A.~L.~Hill}
\affiliation{Harvard-Smithsonian Center for Astrophysics, 60 Garden Street, MS 59, Cambridge, MA 02138, USA}
\author{R.~W.~Mair}
\affiliation{Harvard-Smithsonian Center for Astrophysics, 60 Garden Street, MS 59, Cambridge, MA 02138, USA}
\author{M.~S.~Rosen}
\affiliation{Department of Physics, Harvard University, Cambridge, Massachusetts 02138, USA}
\affiliation{Harvard-Smithsonian Center for Astrophysics, 60 Garden Street, MS 59, Cambridge, MA 02138, USA}
\author{R.~L.~Walsworth}
\affiliation{Department of Physics, Harvard University, Cambridge, Massachusetts 02138, USA}
\affiliation{Harvard-Smithsonian Center for Astrophysics, 60 Garden Street, MS 59, Cambridge, MA 02138, USA}
\author{C.~M.~Marcus}
 \altaffiliation{To whom correspondence should be addressed (marcus@harvard.edu)}
 \affiliation{Department of Physics, Harvard University, Cambridge, Massachusetts 02138, USA}

\date{\today}
\begin{abstract}
Silicon nanoparticles are experimentally investigated as a potential hyperpolarized, targetable MRI imaging agent. Nuclear $T_1$ times at room temperature for a variety of Si nanoparticles are found to be remarkably long ($10^2$ to $10^4\,$s)---roughly consistent with predictions of a core-shell diffusion model---allowing them to be transported, administered and imaged on practical time scales without significant loss of polarization.  We also report surface functionalization of Si nanoparticles, comparable to approaches used in other biologically targeted nanoparticle systems.
 \end{abstract}

\maketitle

The use of nanoparticles for biomedical applications has benefited from rapid progress in nanoscale synthesis of materials with specific optical and magnetic properties, as well as biofunctionalization of surfaces, allowing targeting \cite{Atanasijevic, Akerman,Weissleder:2}, \emph{in-vivo} tracking \cite{Weissleder:2, Gao, Hogemann}, and therapeutic action \cite{Hirsch, Simberg}. For magnetic resonance imaging (MRI), superparamagnetic nanoparticles have extended susceptibility-based contrast agents toward targeted imaging \cite{Weissleder:1}, though achieving high spatial resolution with high contrast remains challenging. An alternative approach is direct MRI of hyperpolarized materials with little or no background signal. Hyperpolarized noble gases\cite{Leawoods,Schroder,Patz} and $\bf ^{13}$C-enhanced biomolecules \cite{Golman:1, Nelson} have demonstrated impressive image contrast, but are limited by short  \emph{in-vivo} enhancement times  ($\sim$~10~s for noble gases \cite{Leawoods}, $\sim$~30~s for $\bf ^{13}$C biomolecules \cite{Golman:1,Nelson}). It is known that bulk silicon can exhibit multi-hour nuclear spin relaxation (\emph{T$_1$}) times at room temperature \cite{Schulman} and that silicon nanoparticles can be hyperpolarized via dynamic nuclear polarization (DNP) \cite{Dementyev}.

Nuclear magnetic resonance (NMR) in silicon has been widely investigated for half a century \cite{Schulman}, and with renewed interest in the context of quantum computation \cite{Ladd}.  The low natural abundance of spin-$1/2$ $^{29}$Si nuclei (4.7\%) embedded in a  lattice of zero-spin $^{28}$Si nuclei isolates the active nuclear spins from one another and from the environment, leading to multi-hour spin relaxation ($T_1$) times, and decoherence ($T_2$) times of tens of seconds \cite{Ladd}.  Moreover, the weak dipole-dipole coupling of the sparse $ ^{29}$Si atoms, together with the isotropic crystal structure and the absence of nuclear electric quadrupole moment conspire to keep any induced nuclear polarization aligned with an external magnetic field, even as the nanoparticle tumbles in space.  This is critical for tracking hyperpolarized nanoparticles in a fluid suspension using MRI, as the nuclear polarization direction can be fixed using a small (mT-scale) applied field.

\begin{figure}[t]
\begin{center}
\includegraphics[width=3.25in]{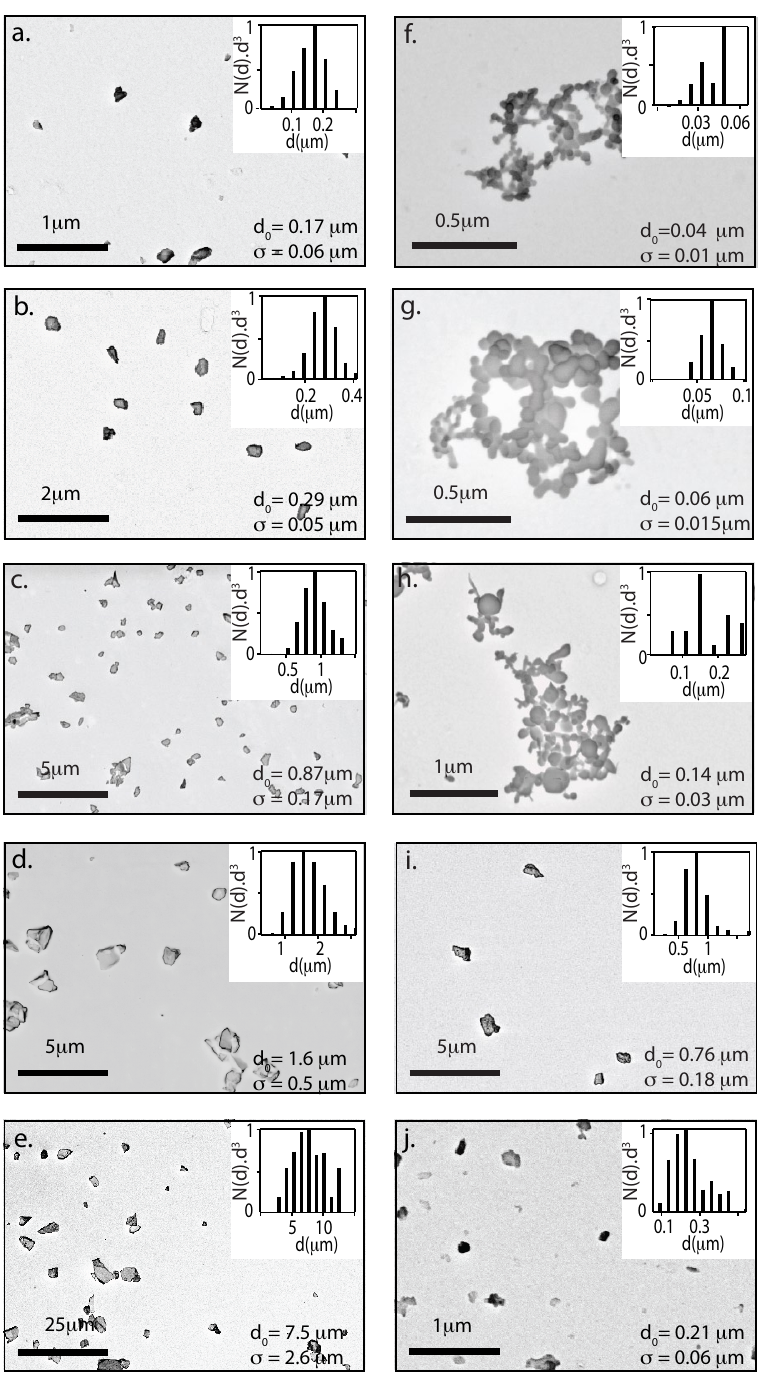}
\caption{Electron micrographs of Si nanoparticles. (a)-(e) ball milling high-resistivity silicon wafer, (f)-(g) wet synthesis (Meliorum), (h) plasma synthesis (MTI), (h) electrical explosion (NanoAmor), (j) ball milling low-resistivity wafer. Insets: Volume-weighted histograms of diameters following size segregation along with averages $d_0$ and standard deviations $\sigma$ based on gaussian fits to distributions.}
\label{fig1}
\end{center}
\end{figure}

Particle size determines regimes of application to biomedicine \cite{Jiang} as well as predicted NMR properties \cite{Dementyev}. Here, we report NMR properties of Si particles spanning four orders of magnitude in mean diameter, from $\sim$40~nm nanoparticles to millimeter-scale granules. We investigate particles made by ball milling high-resistivity (30$-$100~k$\Omega$-cm, residual p-type $\langle 111 \rangle$, Silicon Quest International), and low-resisitivity ($0.01-0.02$~$\Omega$-cm, boron-doped (p-type), $\langle 100 \rangle$ oriented, Virginia Semiconductor) commercial silicon wafers, followed by centrifugal segregation by size \cite{Brown}. We also investigate chemically synthesized Si nanoparticles with mean diameters 40~nm (Meliorum), 60~nm (Meliorum), 140~nm (MTI) and 600~nm (NanoAmor), obtained commercially.   Figure 1 shows representative scanning electron microscope (SEM) images of all measured particles, along with volume-weighted size distributions obtained by SEM image analysis. Dilute suspensions of silicon nanoparticles in ethanol were sonicated for ten minutes before being pipetted onto a vitreous carbon planchett which was mounted on a standard specimen holder with conducting carbon tape. For each sample, $> 1000$ particles were analyzed, sourced from $\sim 50$ images, with particles in contact excluded from the analysis. Particle agglomeration seen in dry Meliorum and MTI samples has been reported in similarly sized silica nanoparticles, but is significantly reduced after pegylation\cite{Xu}. In these cases (Meliorum, MTI), individual measurement of the particle diameter from SEM images was used instead of software analysis.

\begin{figure}[t]
\begin{center}
\includegraphics[width=3.25in]{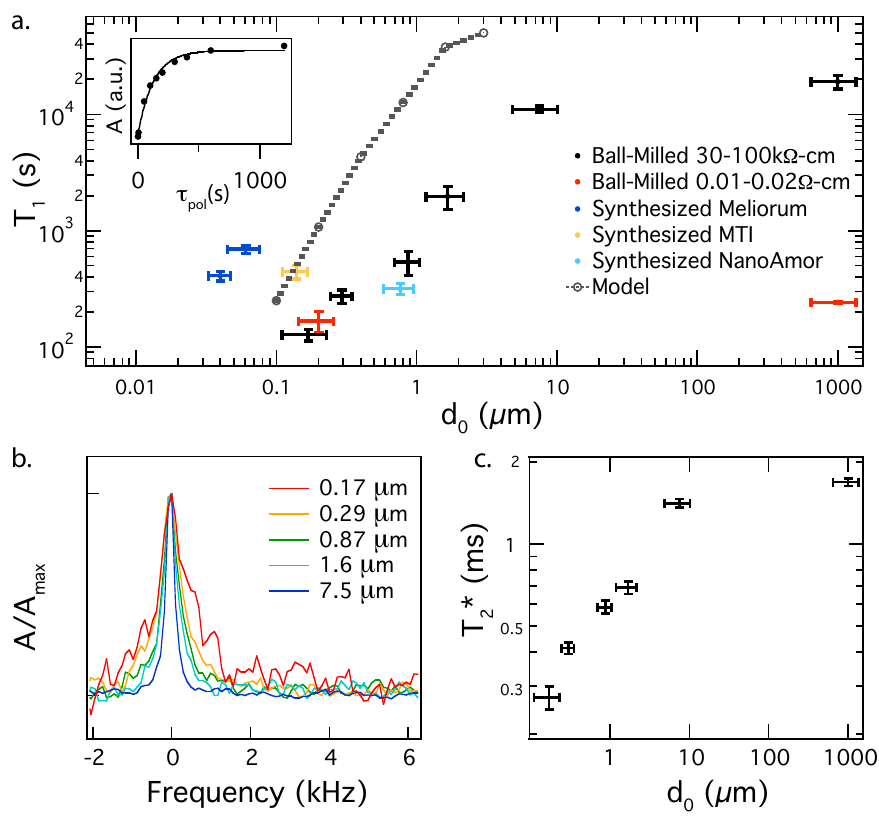}
\caption{(a) Nuclear spin relaxation ($T_1$) times at 2.9 T as a function of particle diameter $d_0$ for various Si particles. Vertical error bars are from exponential fits to relaxation data; horizontal error bars are $\sigma$ of size distributions (see Fig.~1). Also shown is a core-shell diffusion model of $T_1$ dependence on particle diameter, from Ref.~\onlinecite{Dementyev} (dashed curve). Inset: Fourier-transform NMR peak amplitude, A, as a function of polarization time  $\tau_{\rm pol}$ (see text) for the ball-milled high-resistivity particles with $d_0 = 170$~nm. $T_1$ values were measured using a saturation recovery spin echo pulse sequence described in the text.
(b) Amplitude normalized NMR lineshape (A/A$_{\rm max}$) versus frequency for ball-milled high resistivity samples at 4.7~T.
(c) Inhomogeneous dephasing time, $T_2^*$, as a function of mean particle diameter for ball-milled high resistivity samples at 4.7~T.}
\label{fig2}
\end{center}
\end{figure}

Nuclear $T_1$ times of the Si nanoparticles, segregated by size and packed dry in teflon NMR tubes, were measured at room temperature at a magnetic field of 2.9~T using a spin-echo Fourier transform method with a saturation recovery sequence. Following a train of sixteen hard $\pi$/2 pulses to null any initial polarization, the sample was left at field to polarize for a time $\tau_{\rm {pol}}$, followed by a CPMG sequence $(\pi/2)_X$\,-\,$[\tau$\,-\,$(\pi)_Y$\,-\,$\tau$\,-\,echo$]^n$ with $\tau =1$~ms and $n=200$. In Si and other nuclear-dipole-coupled materials echo sequences can yield anomalously long decay tails \cite{Li}.  However, the Fourier amplitude of the echo train still provides a signal proportional to initial polarization\cite{Li}.  Values for $T_1$ are extracted from exponential fits, $A\propto1-e^{-\tau_{\rm {pol}}/T_1}$, to the Fourier amplitude, $A$, of the $n=200$ echoes as a function of polarization time (see Fig.~2a, inset for an example).

Figure 2a shows $T_1$ as a function of (volume-weighted) average particle diameter for the various samples. The high-resistivity ball-milled samples follow a roughly linear dependence on size, $T_1 \propto d_0$, for $d_0 < \sim 10 \mu$m, saturating at  $T_1 \sim 5$~h for larger particles. The trend of increasing $T_1$ in larger particles is qualitatively consistent with a simple shell-core spin diffusion model \cite{Dementyev}(dotted line), which predicts $T_1 \propto d_0^2$ for particles with no internal defects or dopants.  The low-resistivity ball-milled particles have $T_1 \sim 200$~s, independent of size. Smaller commercial particles formed by wet synthesis (Meliorum) and plasma synthesis (MTI) have $T_1$ times as long as 700~s. This is significantly longer than the predicted values, as this model assumes instantaneous spin relaxation at the surface of the particle. Larger commercial particles formed by electrical explosion (NanoAmor) have shorter $T_1$ than the comparably sized high-resistivity ball-milled particles. Powder x-ray diffraction measurements (not shown) indicate that ball milling induces partial polycrystallinity, consistent with previous studies \cite{Shen}.  We speculate that $T_1$ is reduced for the smallest ball-milled particles, compared to the synthesized particles, by coupling of nuclear spins to paramagnetic defects at the interface between crystallites. Electron Spin Resonance (ESR) measurements (see Supplementary Material S1) on ball-milled and synthesized particles show a single peak corresponding to a g-factor of g=2.006, characteristic of P$_b$-type defect centers \cite{Nishi}. For a fixed volume of particles, a larger ESR signal is seen in the smaller samples, suggesting that the surface, rather than the bulk, is the source of paramagnetic defects. A detailed study relating ESR and NMR properties will be reported later.

In addition to noting the remarkably long $T_1$ times for all Si measured particles relative to previously reported hyperpolarized MRI imaging agents\cite{Patz, Leawoods,Golman:1, Nelson}, we also note that $T_1$ in the Si system can be tuned by size and doping, allowing optimization for specific applications, from virtual colonoscopy to molecular imaging. 
 	
Averaged NMR spectra for ball-milled high-resistivity samples of various particle sizes are shown in Fig. 2b. Each spectrum is taken from a series of summed free induction decay traces following polarization for a time 3$T_1$ at a field of 4.7~T corresponding to 39.7~MHz, taken on a Bruker DMX-200 NMR console. In post processing, these spectra have been individually phase adjusted. Whereas $T_1$ changes by two orders of magnitude over the range of measured particle sizes, $T_2^*$ changes only by factor of $\sim$6 over the same range. It is expected that the rapid tumbling of particles in a liquid suspension will increase  $T_2^*$ considerably, restoring resolution. For biologically targeted imaging, the bound particles may not tumble, but may move enough for spectral narrowing to take place. This will be addressed in future work.

\begin{figure}[t]
\begin{center}
\includegraphics[width=2.7in]{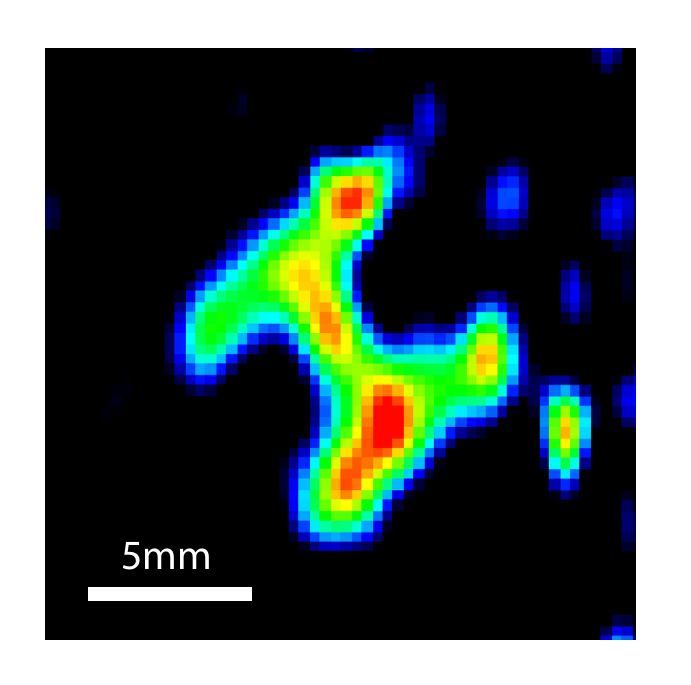}
\caption{Magnetic resonance imaging of hyperpolarized Si nanoparticles. An H-shaped cavity filled with high-resistivity Si particles ($d_0$ = 1.6~$\mu$m) pre-polarized at low temperature (T = 4.2~K) and high magnetic field (B = 5~T) for 60~h and warmed and transferred to a 4.7~T imager. See text for imaging details. No Si image could be obtained without hyperpolarization.}
\label{fig3}
\end{center}
\end{figure}

To demonstrate MRI of hyperpolarized Si nanoparticles, we filled a small cavity (in the shape of the letter H) with high-resistivity ball-milled particles ($d_0$ = 1.6$\mu$m). The sample was left to polarize at low-temperature (4.2~K) and high field (5~T) for 60~h, which enhanced the $^{29}$Si nuclear spin polarization by approximately a factor of 15 over room temperature. We then removed the sample from the polarizing cryostat and transferred it to the 4.7~T Bruker DMX-200 imager with a micro-imaging gradient set, requiring $\sim1$ minute for the transfer, and imaged the phantom using a small tip angle gradient echo sequence \cite{zhao}. Imaging parameters were: tip angle $\theta$ = 9$^{\circ}$,  echo time $\tau$ = 1.2~ms,  field of view = 15~mm, sample thickness = 2.5~cm, single pass (no averaging), acquisition time = 11~s. The resulting image is shown in Fig.~3. We note that much higher image resolution will be possible with DNP polarization \cite{Leawoods,Schroder,Patz,Golman:1,Nelson}.

To examine the applicability of Si nanoparticles to targeted MRI, we prepared the Si nanoparticle surface for attachment to biological-targeting ligands.  Ball-milled high resistivity nanoparticles ($d_0$~=~200~nm) were aminated using either (3-Aminopropyl)triethoxysilane (APTES, Sigma, 99\%) or a 1:2 mixture by volume of APTES with bis-(triethoxysilyl)ethane (BTEOSE, Aldrich, 96\%) or (3-trihydroxysilyl)propyl methylphosphonate (THPMP, Aldrich, 42~wt\% in H$_2$O) (see Fig. 4~a)\cite{Howarter:2}. The surface oxide was first etched with a dilute solution of hydrofluoric acid (8\% in ethanol) followed by resuspension of the particles in ethanol. Approximately 100~mg of silicon nanoparticles were added to 45~mL of acidified 70\% ethanol (0.04\% v/v,  adjusted  to pH~3.5 with HCl) and the solution was placed in an ultrasonic bath for five minutes. Saline (0.15~M) was then added and the solution was shaken for 18-24 hours. Silanes were removed from the nanoparticle solution by washing and resuspending three times in methanol buffer, with the final resuspension performed with 10 mL of methanol buffer.
 Successful amination was assessed using fluorescence spectroscopy (Fig.~4b) using an excitation at 390~nm and emission at 465~nm (SpectraMax Plus, Molecular Devices). The high level of fluorescence observed for aminated particles results from the covalent bonding of surface amino groups with fluorescamine, showing these functional groups were accessible for further reaction.

\begin{figure}[t]
\begin{center}
\includegraphics[width=3.25in]{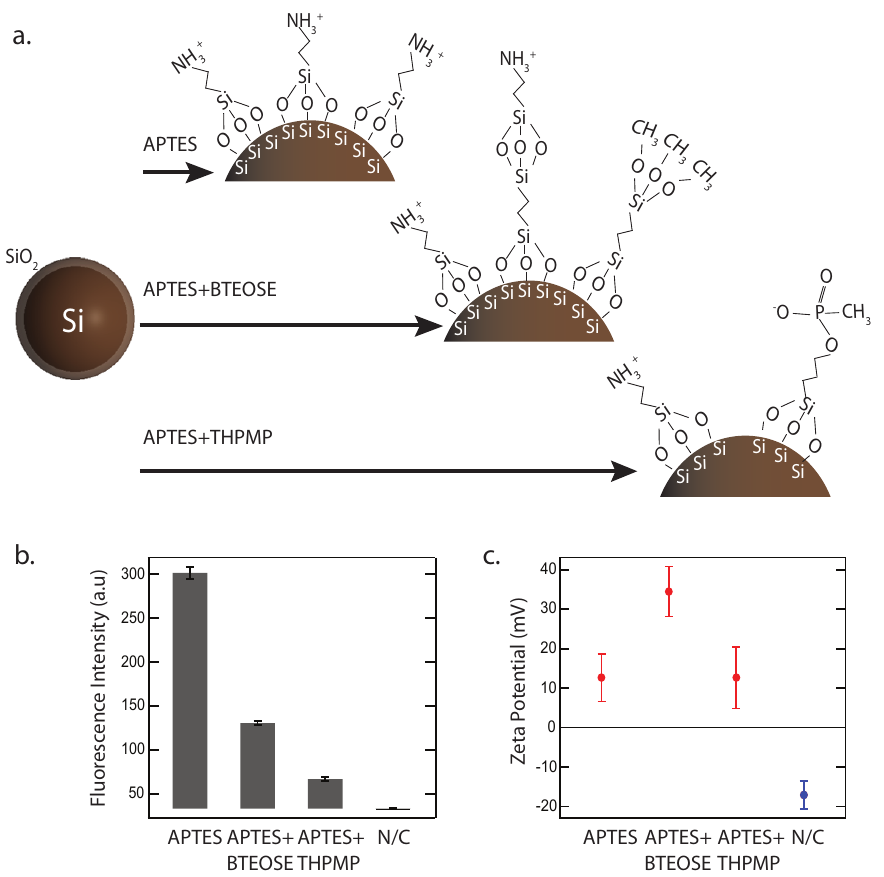}
\caption{Biological surface modification of silicon nanoparticles.  (a) Silicon particles ($d_0$~=~0.2~$\mu$m) were aminated using either (3-Aminopropyl)triethoxysilane (APTES) alone or as a 1:2 mixture by volume of APTES with bis-(triethoxysilyl)ethane (BTEOSE) or (3-trihydroxysilyl)propyl methylphosphonate (THPMP in H$_2$O). (b) Fluorescence spectroscopy confirmed the success of the amination reaction. No fluorescence was evident with the negative control (N/C). (c) A change in the sign of the surface charge, or zeta potential of the particles was evident after amination with the three amine groups when compared to the negative control.}
\label{fig4}
\end{center}
\end{figure}

In addition to chemical assays, the accumulation of amines was indirectly monitored by measuring the particles' surface charge in solution, known as the zeta potential \cite{Jana} (Fig. 4c). The surface of the unmodified silicon nanoparticles is composed of hydroxyl groups from the silicon dioxide and thus shows a negative zeta potential. Particles treated with APTES have surfaces coated with propylamines, which become protonated and positively charged in acidic solutions and show a positive zeta potential \cite{Jana}.
	
Aminated particles were coated with polyethylene glycol (PEG) polymers to confer stability and biocompatibility. PEG coating of silica and iron-oxide nanoparticles has been shown to be non-toxic \cite{Ferrari} and to reduce the rate of clearance by organs such as the liver or kidneys, thus increasing the particle's circulation time \emph{in-vivo} \cite{Ferrari}. Pegylation was performed with either $\alpha$-methyl-PEG-succinimidyl $\alpha$-methylbutanoate (mPEG-SMB) (Nektar) or maleimide-PEG-N-hydroxysuccinimide (MAL-PEG-NHS) (Nektar). Both SMB and NHS are reactive with amines on the particle surface. 10~mg of PEG was mixed in 500~$\mu$L of methanol buffer and heated briefly at $50\,^{\circ}\mathrm{C}$ to dissolve. Approximately 0.1~mg of aminated particles (100~$\mu$L in solution) were added to this solution and it was placed in an ultrasonic bath for 1Ð3 h. To remove the unreacted PEG, samples were centrifuged and resuspended twice in methanol and finally in a phosphate-buffered saline solution (PBS, 0.1~M Na$_2$HPO$_4$, 0.015~M NaCl buffer).

 The stability of nanoparticles in solution was assessed using both dynamic light scattering (DLS) (Nano ZS90, Malvern) as a measure of the particles' hydrodynamic radius, and visual determination of flocculation and sedimentation.  The particles treated with mPEG-SMB and NHS-PEG-MAL were both stable in phosphate-buffered saline (PBS) for a period of two days, with no significant change in the particles' hydrodynamic radius (see Supplementary Information S2).  As a control, mPEG-Amine polymer, which does not contain amine-reactive groups, was used. The aminated particles treated with mPEG-Amine aggregated after centrifugation and resuspension in PBS. These results are consistent with other reports of the successful pegylation of SiO$_2$ nanoparticles \cite{Xu,Zhang}.

In conclusion, we demonstrate that Si nanoparticles show promise as biologically targeted MRI imaging agents based on their exceptional NMR properties, including their receptivity to hyperpolarization and long nuclear relaxation ($T_1$) times, in the range of minutes to hours.  We investigated Si $T_1$ times as a function of nanoparticle size, dopant concentration and synthesis method. Furthermore, we have demonstrated techniques for fabricating, size-separating and coating Si nanoparticles to satisfy a broad spectrum of design criteria.  Future developments in the chemical synthesis of larger, mono-disperse single-crystal silicon nanoparticles may provide even longer $T_1$ times.  We note that Si nanoparticles may be combined with other material components to provide MRI tracking of the delivery of drugs \cite{Jain} or as a therapeutic agent that allows simultaneous MRI tracking.  The addition of APTES and PEG to the surface of these nanoparticles is a critical step for further surface functionalization and, ultimately, biological targeting.

\begin{acknowledgments}

We thank D. C. Bell, F. Keummeth, F. Kosar, C. Lara, D. Reeves, S. Rodriques, and J. R. Williams for technical contributions and D. J. Reilly, C. Farrar, and B. Rosen for valuable discussions. This work was supported by the NIH under grant no. 1 R21 EB007486-01A1, U54 CA119335, R01 CA124427 and the Harvard NSEC. Part of this work was performed at the Harvard Center for Nanoscale Systems (CNS), a member of the National Nanotechnology Infrastructure Network (NNIN), which is supported by the National Science Foundation under NSF award no. ECS-0335765.
\end{acknowledgments}

\end{document}